\documentclass[pre,a4paper,twocolumn,showpacs,superscriptaddress,floatfix]{revtex4}
\usepackage{amsmath}
\usepackage{amsfonts}
\usepackage{amssymb}
\usepackage{bm}
\usepackage{graphicx}
\usepackage{hyperref}

\begin{document}

\title{Anomalous translational velocity of vortex ring with finite-amplitude 
Kelvin waves}

\author{C.~F.~Barenghi}
\affiliation{School of Mathematics, University of Newcastle,
Newcastle NE1 7RU, UK}

\author{R.~H\"anninen}
\affiliation{Department of Physics, Osaka City University, 
Sugimoto 3-3-138, 558-8585 Osaka, Japan}

\author{M.~Tsubota}
\affiliation{Department of Physics, Osaka City University, 
Sugimoto 3-3-138, 558-8585 Osaka, Japan}

\date{\today}
\begin{abstract}
We consider finite-amplitude Kelvin waves on an inviscid vortex assuming 
that the vortex core has infinitesimal thickness. By numerically solving 
the governing Biot-Savart equation of motion, we study how the frequency 
of the Kelvin waves and the velocity of the perturbed ring depend on the
Kelvin wave amplitude. In particular, we show that, if the amplitude of 
the Kelvin waves is sufficiently large, the perturbed vortex ring moves 
backwards.
\end{abstract}
%
\pacs{47.32.cf, 67.40.Vs, 67.57.Fg}
\maketitle

\section{Introduction}
 
Vortex rings are among the most important and most studied objects of 
fluid mechanics \cite{Sharif,Saffman}. It has been known since the times 
of Lord Kelvin \cite{Kelvin} that a vortex ring is subject to wavy 
distortions (sinusoidal displacements of the vortex core) called Kelvin 
waves \cite{Maxworthy}. In the case of viscous vortex rings, the stability 
of these waves is a problem with subtle aspects \cite{stability} which are 
still the focus of intense mathematical scrutiny \cite{Fukumoto}. Our 
concern is the simpler case in which the fluid is inviscid and the vortex 
core has infinitesimal thickness. This case refers to the idealized context 
of classical Euler fluids, but is realistic for superfluids, which have 
zero viscosity and microscopic vortex core thickness.

Vortex rings have indeed been central to superfluidity \cite{Donnelly} 
since the pioneering experiments on the nucleation of quantized vorticity 
by moving ions \cite{ions}, the early investigations into rotons as ghosts 
of vanished vortex rings \cite{ghost}, and the nature of the superfluid 
transition \cite{lambda}. The current interest in superfluid vortex rings 
extends to the physics of cold atomic gases \cite{BEC} and the discovery 
of new nonlinear solutions \cite{NLSE} of the Gross-Pitaevskii's nonlinear 
Schr\"odinger equation (NLSE) for a Bose-Einstein condensate. Vortex rings 
are also important in the study of superfluid turbulence \cite{Vinen}. For 
example, they have been used as tools to study the Kelvin wave cascade 
\cite{Kivotides} which is responsible for the dissipation of turbulent 
kinetic energy near absolute zero, and to investigate the effects of vortex 
reconnections \cite{reconnections}, which are the key feature of turbulence; 
they are also used as simple models of the vortex loops which make up the 
turbulence \cite{Tsubota}. 

Kelvin waves play a role in all examples listed above. The dispersion 
relation of Kelvin waves of infinitesimal amplitude $A$ on a circular vortex 
ring of given radius $R$, circulation $\kappa$, and vortex core radius 
$a$ is \cite{disprel}

\begin{equation}
\omega = 
\frac{\kappa}{2\pi a^2}\left(1-\sqrt{1+ka\frac{K_0(ka)}{K_1(ka)}}\right),
\label{e.disprel}
\end{equation} 

\noindent

\noindent
where $\omega$ is the angular velocity of the wave and $k$ the wave number.
Functions $K_n(x)$ are modified Bessel functions of order $n$. The above
dispersion relation is also valid for waves on a straight vortex 
\cite{disprelStraighV}. The properties of small-amplitude Kelvin waves have 
been already investigated \cite{BDV}, but little is known of what happens at 
large wave amplitude. The stability problem becomes nonlinear, hence more 
difficult, and a numerical approach is necessary. 

Recently, an astonishing prediction was made by Kiknadze and Mamaladze 
\cite{Mamaladze} that, at sufficiently large amplitude, the perturbed vortex 
ring moves backwards. Unfortunately the prediction arises from numerical 
analysis based on the local induction approximation (LIA) to the exact 
equation of motion, which is the Biot-Savart law (BSL). The advantage of 
the LIA over the BSL is that it is analytically simpler and computationally 
cheaper. If $N$ is the number of discretization points along a vortex 
filament, the cost of the computation grows as $N$ under the LIA, whereas 
under the BSL it grows as $N^2$. The use of the LIA was pioneered by Schwarz 
\cite{Schwarz} in his numerical studies of homogeneous isotropic turbulence. 
His results obtained using the LIA compared reasonably well with results 
obtained using the BSL, because long-range effects tend to cancel out in 
the isotropy vortex configurations which he considered. In less isotropic 
cases, however, for example in rotating turbulence \cite{rotating-turbo}, 
the LIA may not be a good approximation. In particular, the LIA yields 
wrong predictions about the stability and motion of vortex knots 
\cite{Ricca}, structures which are geometrically similar to (although 
topological different from) the perturbed vortex rings considered by 
Kiknadze and Mamaladze \cite{Mamaladze}.

Our first aim is thus to use the exact BSL to investigate the claim of 
Kiknadze and Mamaladze that the perturbed vortex ring can move backwards 
\cite{Mamaladze}. Our second aim is to carry out a more detailed examination 
of the effects of large-amplitude Kelvin waves on the motion of a vortex 
ring.

\section{Model}

Our approach is based on the vortex filament model of Schwarz \cite{Schwarz}
which is appropriate to superfluid helium due to the smallness of the vortex 
core radius $a$ compared to the radius of the vortex ring $R$. Essentially, 
a vortex is treated as a topological line defect, that is to say a curve in 
three-dimensional space. In the absence of dissipation (zero temperature),
the vortex at the point ${\bf r}$ moves with velocity $d{\bf r}/dt={\bf v}_L$
where ${\bf v}_L$ is equal to the local superfluid velocity ${\bf v}_s$ that 
is given by the following Biot--Savart line integral calculated along the 
entire vortex configuration:

\begin{equation}
{\bf v}_{s}({\bf r},t) = \frac{\kappa}{4\pi}
\oint\frac{({\bf s}-{\bf r})\times d{\bf s}}
{\vert {\bf s}-{\bf r}\vert^3}.
\label{e.bs}
\end{equation}

\noindent
Here ${\bf s}$ denotes a variable location along the vortex filament. To 
implement the BSL, the vortex configuration is discretized into a large 
number of segments. The technique to handle the singularity that one meets 
when one tries to evaluate the integral at those discrete points that are 
used to describe the vortex line can be avoided by splitting the integral 
into local and nonlocal parts \cite{Schwarz}. The velocity of a point 
${\bf s}$ on the vortex is thus

\begin{equation}
{\bf v}_L = \frac{\kappa}{4\pi}{\bf s}'\times {\bf s}'' 
\ln\left(\frac{2\sqrt{l_{+}l_{-}}}{e^{1/2}a}\right)
+\frac{\kappa}{4\pi}\oint^{'}\frac{({\bf s}_1-{\bf s})\times d{\bf s}_1}
{\vert {\bf s}_1-{\bf s}\vert^3}. 
\label{e.vs}
\end{equation}

\noindent
where $\xi$ is the arc length, the vectors ${\bf s}'=d{\bf s}/d\xi$,
${\bf s}''=d^2{\bf s}/d\xi^2$ are, respectively, the local tangent and the 
local normal to the vortex at the point ${\bf s}$. The quantities $l_{-}$ and 
$l_{+}$ are the lengths of the line segments connected to the discretization 
point ${\bf s}$ and the prime above the integral symbol means that the line
integration now extends only along the remaining vortex segments. One should
note that we use a hollow core vortex, which results that the scaling factor
in front of $a$ in Eq.~(\ref{e.vs}) is $\exp(1/2)$ rather than $\exp(1/4)$ 
which is for a solid rotating core and appears in a paper by Schwarz 
\cite{Schwarz}. The recent progress in using the Gross-Pitaevskii nonlinear 
Schr\"odinger equation for quantum fluids suggests that the hollow core model 
should be more appropriate \cite{GP}.
The exact value of the core size is not important here. What matters is that 
it is orders of magnitudes smaller than the radius of the ring or the amplitude 
of the waves, so that we can use the concept of vortex filament. For example,
in a typical helium turbulence experiment the measured vortex line density $L$ 
is $10^4$ or $10^6$~cm$^{-2}$, which means that the intervortex spacing is 
$1/\sqrt{L}$ = 0.01 or 0.001 cm, which is a million or hundred thousands 
times bigger than the vortex core radius ($10^{-8}$~cm) in $^4$He.

The local induction approximation (LIA) is obtained by neglecting the 
nonlocal part and is typically written in the form:
\begin{equation}
{\bf v}_L = \beta{\bf s}'\times {\bf s}'' , 
\label{e.lia}\end{equation} 

\noindent
where $\beta = \kappa\ln(c\langle{R}\rangle/a)/4\pi$, $\langle{R}\rangle$ is 
some average curvature, and $c$ is of order unit; the last two parameters are 
adjusted to obtain better agreement with full nonlocal calculations. By 
choosing $c$ = $8\exp(-1/2)$ and $\langle{R}\rangle$ to be the local radius 
of curvature one obtains fairly good results and additionally a limit that 
gives correctly the velocity for the perfect ring. 

The calculation of the kinetic energy $E$ of the vortex would not be 
accurate if carried out on a three-dimensional mesh around the vortex due 
to rapid changes of the velocity field near the vortex core. Fortunately in 
our case the vortex filament forms a closed loop and the velocity field goes 
to zero at infinity (the calculation is performed in an infinite box), hence 
it is appropriate \cite{caveat} to use Saffman's formula \cite{Saffman}
\begin{equation}
E = \kappa\rho_s\oint {\bf v}_s\cdot{\bf s}\times d{\bf s},
\label{e.enemom}\end{equation}

\noindent
where the line integration is performed along the vortex filament and $\rho_s$ 
is the superfluid density. 

The initial condition consists of a vortex ring of radius $R$ with superimposed 
$N$ Kelvin waves of amplitude $A$ (that is, the wavelength of the perturbation 
is $2 \pi R/N$). Using cylindrical coordinates $r$, $\phi$, and $z$, the 
Cartesian coordinates of the initial vortex ring are thus
\begin{eqnarray}\label{e.initconf}
  x &=& R\cos\phi+A\cos(N\phi)\cos\phi, \nonumber \\
  y &=& R\sin\phi+A\cos(N\phi)\sin\phi, \\
  z &=& -A\sin(N\phi) .\nonumber 
\end{eqnarray}

\noindent
In the absence of Kelvin waves ($A=0$) the circular vortex ring moves in the
positive $z$ direction with self-induced translational speed \cite{veloring} 
\begin{equation}\label{e.translational}
v_{\rm ring}=\frac{\kappa}{4\pi R}[\ln{(8R/a)} - 1/2].
\end{equation}
\noindent
We have tested that, in the case of a circular ring, our numerical
method agrees fairly well with this result.

\noindent
All results presented here are obtained using ring radius $R = 0.1$ cm 
and values of $a$ and $\kappa$ which refer to $^4$He
($\kappa = h/m_4 = 9.97 \times 10^{-4}$ cm$^2/{\rm s}$, where $m_4$ is the 
mass of one atom, and $a=1.0\times 10^{-8}$ cm). The dependence of the 
results on $a$ is small, since $a$ appears only in the slow varying 
logarithmic term in Eq.~(\ref{e.vs}). 

The numerical method to evolve the perturbed vortex ring under the BSL is 
based on a fourth-order Runge-Kutta scheme. The spatial discretization 
is typically $\Delta \xi/R=0.02$ and the time step 
$\Delta{t}=0.5 \times 10^{-3}$ s. The time step is well below the one 
that for a given space resolution provides stable motion of a circular 
vortex ring without fluctuations and resolves the oscillations of the 
Kelvin waves. Numerical calculations are also performed using the LIA to 
compare against the exact BSL.

We are unable to perform a precise stability analysis of large-amplitude 
Kelvin waves under the Biot-Savart Law or a stability analysis of the 
Runge-Kutta scheme when applied to the Biot-Savart motion -- both problems 
are practically impossible. We find that for very large times (larger then 
reported in the following section) the perturbed vortex ring always breaks 
up at some point (that is, first deforms and later possibly attempts to 
reconnect with itself). We do not know whether this fate indicates an 
instability of the vortex for large-amplitude Kelvin waves or a numerical 
instability. What matters is that the lifetime of the perturbed vortex and 
the spatial range that it travels are much larger than the time scale of the 
Kelvin oscillations and the size of the ring itself, because it implies that 
the results which we describe are physically significant and observable in 
a real system.

\section{results}

\begin{figure}[tb]
\centerline{\includegraphics[width=0.95\linewidth]{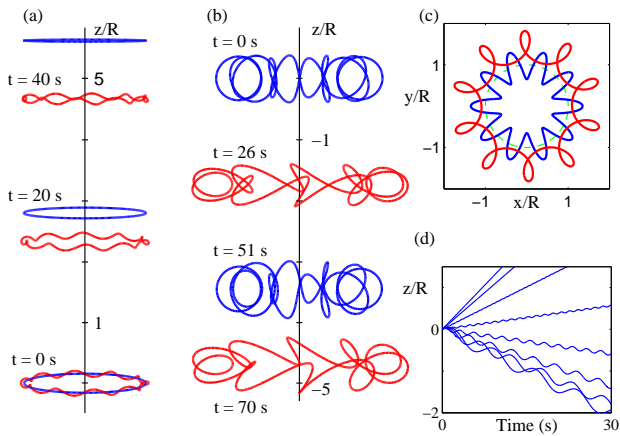}}
\caption{(Color online) Snapshots of the vortex ring of radius $R = 0.1$ cm 
perturbed by $N=10$ Kelvin waves of various amplitude $A$ taken during the 
motion of the vortex. In the {\it left panel} (a) the amplitude of the Kelvin 
waves is small, $A/R = 0.05$, but the perturbed vortex ring (red color) already 
moves slower than the unperturbed vortex (blue color). In the {\it center panel} 
(b) the Kelvin waves have large amplitude, $A/R$ = 0.35, and the perturbed 
vortex ring moves backwards (negative $z$ direction) on average. The 
{\it top right panel} (c) shows the top ($xy$) view of the large amplitude 
vortex at $t = 0$ s (blue) and $t = 26$ s (red, outermost). For 
comparison, a nondisturbed vortex is shown with dashed line (green). The 
{\it lower right panel} (d) gives the averaged location of the ring as a 
function of time. From top to bottom the curves correspond to $A/R$ = 0.0, 
0.05, 0.10, \ldots, 0.35.}
\label{f.fig1}
\end{figure}

\begin{figure}[tb]
\centerline{\includegraphics[width=0.95\linewidth]{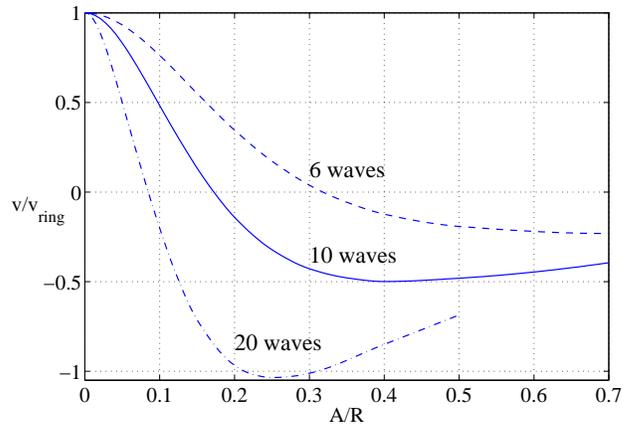}}
\caption{Average translational velocity of the vortex ring as a function of 
the initial oscillation amplitude $A/R$. Velocity is scaled by the velocity 
of the unperturbed ring, $v_{\rm ring}$. The dash-dotted line corresponds 
to $N$ = 20, solid line to $N$ = 10, and the dashed line to $N$ = 6 in 
Eq.~(\ref{e.initconf}). Critical amplitudes, above which the velocities become 
negative, are $A/R$ = 0.085, 0.17, and 0.32, respectively.}
\label{f.fig2}
\end{figure}

The first result of our numerical simulations is that Kiknadze and 
Mamaladze's prediction \cite{Mamaladze} obtained using the LIA is indeed 
correct. Integration of the motion using the exact BSL shows that, provided 
the amplitude of the Kelvin waves is large enough, the vortex ring moves 
(on the average) backwards.  This result is illustrated in Figs.~\ref{f.fig1} 
and \ref{f.fig2}: the former shows snapshots of the ring at different times 
as it travels, the latter gives the average translational velocity of the 
ring along the $z$ direction as a function of the amplitude $A$ of the Kelvin 
waves. It is apparent that the translational velocity decreases with 
increasing amplitude of the Kelvin waves and can even become negative. 

At some critical value of the amplitude $A$ the translational velocity is zero 
and the perturbed vortex ring hovers like a stationary helicopter. In the case 
of $N = 10$ Kelvin waves this happens when $A/R = 0.17$ approximately,
which is quite close to the LIA prediction, $A/R = 0.16$. For 
$N = 6$ and $N=20$ the critical value is, respectively, $A/R$ = 0.32 and 
$A/R = 0.085$. This dependence of the critical amplitude on $N$ is in 
approximate agreement with the LIA prediction \cite{Mamaladze}.

The backward velocity of the perturbed vortex ring depends nonlinearly on 
the amplitude $A$ of the Kelvin waves. At large enough amplitude $A$ this 
velocity will slow down. This can be clearly seen in Fig.~\ref{f.fig2}. 
The Kelvin waves, that can be imagined to behave like small vortex rings, 
tend to turn backwards, or more precisely, on the direction opposite to 
the motion of the unperturbed vortex ring. The larger the amplitude the 
larger fraction of the ring velocity is oriented downwards. This is 
compensated by the decrease in velocity of the single ring, which is 
inversely proportional to the amplitude, resulting an optimum value at 
some amplitude. For $N$ = 20 the optimum amplitude $A \approx 0.25R$ 
resulting a downward velocity that is already slightly higher than the 
velocity upwards of the unperturbed ring. 

In addition to Kelvin waves, the translational velocity of the vortex 
ring can be reduced by having an additional swirl velocity along the 
vortex core. This was considered by Widnall, Bliss, and Zalay 
\cite{Widnall}. However, this effect does not matter in our limit of 
thin-core vortices, which is relevant to superfluids.

The dispersion relation of large-amplitude Kelvin waves can be obtained 
by tracking the motion of the vortex on the $y=0$ plane, for example. If 
the amplitude $A$ of the Kelvin wave is small, the vortex draws a circle 
at approximately the same angular frequency that is obtained analytically 
for small-amplitude Kelvin waves and given by Eq.~(\ref{e.disprel}). In 
the long wavelength limit ($k \rightarrow 0$) this relation becomes 

\begin{equation}
\omega = -\frac{\kappa k^2}{4\pi}\left[\ln\left(\frac{2}{ka}\right) 
-\gamma\right],
\end{equation} 

\noindent
where $\gamma = 0.5772\cdots$ is Euler's constant and the negative sign 
only indicates that the Kelvin waves rotate opposite to the circulation. 
Again the above equation differs slightly ($-\gamma$ in stead of 
$1/4-\gamma$) from the form given by Schwarz \cite{Schwarz}, but this is 
again only due to the definition of the core type.

\begin{figure}[tb]
\centerline{\includegraphics[width=0.95\linewidth]{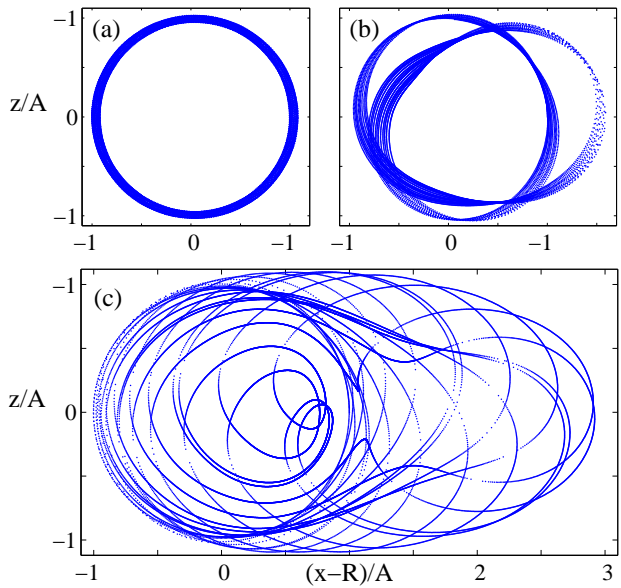}}
\caption{Curve drawn by the vortex at $y$ = 0 plane. Here the $z$ coordinate 
is the coordinate relative to the average location of the vortex and $N$ = 10 
in Eq.~[\ref{e.initconf}]. In the {\it top left panel} (a) the amplitude is 
$A/R$ = 0.05 and in the {\it top right panel} (b) $A/R$ = 0.20. In both panels 
only the first 30 sec are shown. The thickness of the plotted curve arises 
from the chaotic motion rather than initial transient. The {\it bottom panel} 
(c) corresponds to $A/R$ = 0.50 and we have drawn the curve for the first 90 
sec. The time step between the markers is 2 ms; it is apparent that at large 
amplitudes the vortex is far from a sinusoidal helix and that the rotational 
speed at $y$ = 0 plane varies significantly.}
\label{f.fig3}
\end{figure}

\begin{figure}[tb]
\centerline{\includegraphics[width=0.95\linewidth]{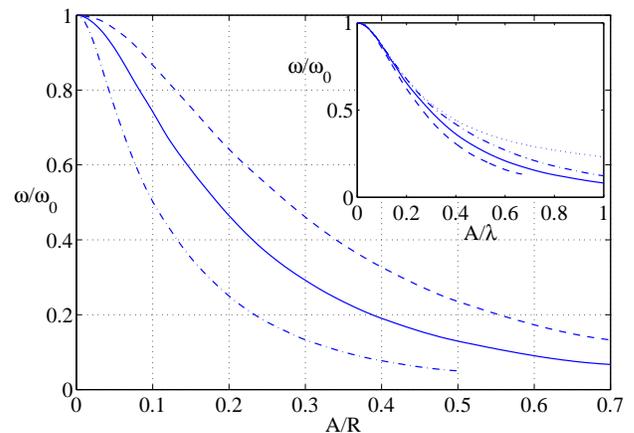}}
\caption{{\it Main figure:} Angular frequency of Kelvin waves $\omega$ 
relative to the value $\omega_0$ obtained in the small amplitude limit 
$A/R=0.001$ and presented as a function of the wave amplitude $A/R$. The 
dashed line is for $N = 6$, the solid line for $N = 10$, and the dash-dotted 
line for $N$ = 20. {\it The inset} shows the same when plotted as a function 
of $A/\lambda$, where $\lambda$ is the wavelength of the Kelvin wave. The 
additional dotted line is the result obtained for straight vortex when using 
a wavelength of 0.1~cm together with periodic boundary conditions and using 
25 periods above and below to numerically determine the vortex motion.}
\label{f.fig4}
\end{figure}

We find that if we increase the amplitude of the Kelvin waves on the
ring then the angular frequency decreases, a result which we also verified
in the case of a straight vortex. Some example curves drawn by the vortex on 
the $y=0$ plane are shown in Fig.~\ref{f.fig3}. The average angular frequency 
is plotted in Fig.~\ref{f.fig4}, which shows also the dispersion relation of 
waves on a straight vortex for comparison. 

\begin{figure}[htb]
\centerline{\includegraphics[width=0.95\linewidth]{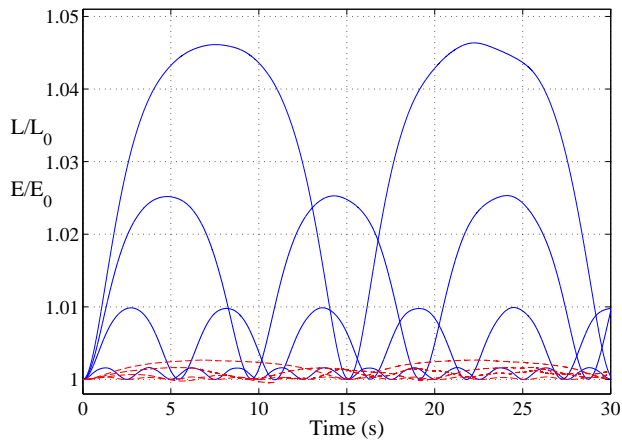}}
\caption{(Color online) The observed vortex length compared with the initial 
length $L_0=2\pi\sqrt{R^2+N^2 A^2}$ is illustrated by solid (blue) lines and 
plotted as function of time in case of $N$=10. For comparison, the dashed 
(red) lines show the fluctuations in energy that are due to numerical errors 
and which can be reduced by increasing the space resolution. With increasing 
amplitude of oscillations the parameters for $A/R$ shown are: 0.20, 0.30, 
0.40, and 0.50.}
\label{f.fig5}
\end{figure}

It is important to notice that, under the LIA used by Kiknadze and 
Mamaladze \cite{Mamaladze} the vortex length remains constant \cite{Schwarz}, 
whereas the quantity which is conserved under the exact BSL is the energy. 
Length and energy are proportional to each other only if the vortex filament
is straight, which is not the case in our problem. Indeed, further 
investigation reveals that the vortex motion contains two characteristic 
frequencies. The first is the Kelvin frequency and the second is the 
frequency that is related to the oscillations of the vortex length and 
illustrated in Fig.~\ref{f.fig5}. If the ratio of the two periods is rational 
one observes a fully periodic motion (in addition to translational motion 
along the $z$ axis). At some values of the amplitudes which we calculated, 
this condition is almost satisfied. At higher values of amplitude one observes 
that the average radius of the vortex ring oscillates, as shown in 
Fig.~\ref{f.fig1}. These variations in the total length were observed but 
not discussed in a recent calculation of the motion of vortex rings using 
the NLSE model \cite{Leadbeater}.

The accuracy of our numerical method is tested by calculating the energy of the 
vortex ring. At zero temperature, without any dissipation,  the energy (and the 
momentum) should remain constant. This condition can be quite well satisfied in 
our calculations. We do get some small oscillations in energy, as seen in 
Fig.~\ref{f.fig5}, but we have checked that by increasing the space resolution 
we can reduce them at will, whereas the oscillations in length are independent 
of the numerical resolution.

\section{Conclusion}
It is well known that a circular vortex ring has a translational velocity 
which arises from its own curvature (the smaller the radius $R$ of the ring, 
the faster the ring travels). Using the exact Biot-Savart law, we have 
analyzed the motion of a vortex ring perturbed by Kelvin waves of finite 
amplitude. We have found that the translational velocity of the perturbed 
ring decreases with increasing amplitude; at some critical amplitude the 
velocity becomes zero, that is, the vortex ring hovers like a helicopter.
A further increase of the amplitude changes the sign of the translational
velocity, that is, the vortex ring moves backward. Our finding confirms 
preliminary results obtained by Kiknadze and Mamaladze using the local 
induction approximation \cite{Mamaladze}.

This remarkable effect is due to the tilt of the plane of the Kelvin waves 
which induce motion in the ``wrong'' direction. The magnitude of the tilt 
oscillates and what results is a wobbly translational motion in the backward 
direction. We have also found that the frequency of the Kelvin wave decreases
with increasing amplitude and that the total length of the perturbed vortex 
ring oscillates with time. This oscillation in vortex length is related to 
the oscillation of the tilt angle.

Time of flight measurements of large, electrically charged, perturbed vortex 
rings in $^4$He could easily detect the decreased translational velocity. 
Another context in which the effect can be studied is Bose-Einstein 
condensation in ultra-cold atomic gases, which allow simple visualization of 
individual vortex structures. For these systems, however, it would be necessary 
to assess the effect of the nonhomogeneity of the superfluid.

\section{Acknowledgments}
The research of C.F.B was supported by EPSRC Grant No. GR/T08876/01 and Grant
No. EP/D040892/1. M.T. acknowledges the support of a Grant-in-Aid for Scientific Research
from JSPS (Grant No. 18340109) and a Grant-in-Aid for Scientific Research
on Priority Areas from MEXT (Grant No. 17071008).




\begin{thebibliography}{99}
\bibitem{caveat}
The energy of a circular vortex ring computed according to this formula slightly 
differs from the value $E=\rho_s \kappa^2 R[\ln{(8R/a)}-3/2]/2$ \cite{BDV} but 
we neglect the small discrepancy which appears in a logarithmic term and depends 
on slightly different assumptions about the core structure.
\end{thebibliography}
\end{document}